\documentclass[12pt,preprint]{aastex}

\shorttitle{Thermal Afterglows of Neutron Stars}
\shortauthors{Hui et al.}

\begin{document}

\title{The Effects of Rotation on Thermal X-ray Afterglows Resulting from Pulsar Glitches}

\author{C. Y. Hui\altaffilmark{1} and K. S. Cheng\altaffilmark{2}}
\affil{Department of Physics, The University of Hong Kong, PRC}

\altaffiltext{1}{e-mail address: h9840329@hkusua.hku.hk}
\altaffiltext{2}{e-mail address: hrspksc@hkucc.hku.hk}

\begin{abstract}
We derive the anisotropic heat transport equation for rotating
neutron stars and the thermal equilibrium condition in a
relativistic rotating axisymmetric star is also derived through a
simple variational argument. With a simple model of neutron star,
we model the propagation of heat pulses resulting from transient
energy releases inside the star. {\bf Such sudden energy release
can occur in pulsars during glitches.} Even in slow rotation limit
($\Omega\leq 1\times 10^{3}\mbox{s}^{-1}$), the results with
rotational effects involved could be noticeably different from
those obtained with a spherically symmetric metric in terms of the
time scales and magnitudes of thermal afterglow. The effects of
gravitational lensing and frame dragging on the X-ray light curve
pulsations are also studied. The results we obtain indicate that
the effect of rotation on the light curve modulation is small, and
the spacetime curvature effect predominates. The metric components
and rotational deformation on the stellar structure are calculated
by using Hartle-Thorne formalism. {\bf We have applied our model
to study the thermal response time scales of pulsars after
glitches and find that the centrifugal force produced by rotation
can substantially reduce the response time by a factor of three
between non-rotating star and rotating star with $\Omega \sim 900
s^{-1}$. The equation of state can also affect the duration of
response.}
\end{abstract}


\section{Introduction}

The rotation-powered pulsars will spin down as they radiate. During the spinning-down epoch, some pulsars exhibit
dramatic events call glitches, which is a sudden spinning-up of the star. Many observations have identified such
phenomena (Krawczyk et al. 2003; Hobbs et al. 2002; Lyne, Shemar, \& Smith 2000; Wang et al. 2000).
There are numerous of mechanism have been proposed to explain the origin of pulsar glitches, while the
superfluid-driven mechanism (Anderson \& Itoh 1975; Alpar et al. 1984) and the starquake mechanism
(Ruderman 1969; Baym \& Pines 1971) are most widely accepted.
The starquake mechanism is based on the idea that a neutron star possesses a solid crust. As the
star spins down, centrifugal force on the crust decreases and gravity pulls the crust to a
less oblate equilibrium shape. This change in stellar shape induces stress in the crust.
However, the rigidity of the solid crust resists this stress and the shape remains more oblate than the equilibrium
value. When the crust stresses reach a critical value, the crust cracks and the glitch energy is
released in a small volume at the weak regions in the solid crust. This localized energy releasing induces an
uneven heating of the surface which corresponds to the `spot' case. It should be noted that the
starquake-driven glitches can occur anywhere within the crust where Coulomb lattice exists.\\
\indent The superfluid-driven mechanism (Anderson \& Itoh 1975) suggests that the spin-up of the
crust is produced by transferring angular momentum from a rotating superfluid to the more slowly rotating crust.
Apart from spinning up, the mechanism also produces frictional heating and hence the local energy dissipation
(Alpar et al. 1984; Shibazaki \& Lamb 1988). Moreover, critical angular speed difference for
spinning is inversely proportional to the distance from the rotational axis, so only equatorial region
shows concern to this process and hence a ring structure at the rotational equator is produced (Cheng et al. 1988).
It should be noted that the superfluid-driven glitches can occur only in the inner crust where superfluid and
Coulomb lattices coexist.\\
\indent Some other glitch mechanisms are proposed. For example, Link \& Epstein (1996) have proposed
a thermal glitch mechanism. In this model, a large increase in the vortex creep rate is induced by a
temperature perturbation. As a result, the superfluid quickly loses angular momentum and delivers a
spin-up torque to the crust.
Carter, Langlois \& Sedrakian (2000) have suggested that centrifugal buoyancy forces are the origin
of pressure gradients sufficient to crack the crust and allow outward vortex motion.\\
\indent Serveral authors (Van Riper et al. 1991; Chong \& Cheng 1994; Hirano et al. 1997;
Cheng, Li, \& Suen 1998; Tang \& Cheng 2001) have suggested that the thermal evolution of
a pulsar after glitches would provide a good method to determine the equation of state for neutron stars.
The glitch energy was first assumed to be released in a spherical shell at a certain density inside the pulsar (
Van Riper et al. 1991; Chong \& Cheng 1994; Hirano et al. 1997), though this does not seem to be realistic.
They also show that the glitches in the `shell'
case cannot produce very significant observed results for young pulsars.\\
\indent Cheng, Li, \& Suen (1998) argue that if a good fraction of
glitch energy is released in a small volume, namely, the `spot'
case, then instead of heating up the entire stellar surface, even
a small fraction of glitch energy can heat up a small area of the
stellar surface drastically. This would result in a periodic hard
thermal X-ray pulse emission that should stand out clearly from
the soft X-ray background. {\bf They suggest that by observing the
time scales of thermal X-ray afterglows resulting from glitches, the equations of state of neutron stars
can be determined.} However, the energy transport in the `spot'
case is clearly not spherically symmetric. They derive the
general expression of the relativistic thermal transport and
energy balance equations without assuming spherical symmetry and use these
equations to study the evolution of the hot spot on the surface of
neutron star after glitches.\\
\indent Tang \& Cheng (2001) incorporate the relativistic light bending effect (Pechenick, Ftaclas, \& Cohen 1983) and magnetic field effect (Page 1995) in their calculations. They show that these effects can significantly affect the intensity and the pulse shape of the transient X-rays
resulting from glitches.\\
\indent Apart from the equation of state, glitch mechanism also affects the thermal responses. Larson \& Link (2002) simulate the emergence of thermal wave at the stellar surface with two different glitch models and compare the results with the data from Chandra observations of thermal emission from the 2000 January glitch in the Vela pulsar (Helfand, Gotthelf \& Halpern 2000).\\
\indent All the aforementioned studies have not considered the effects of rotating metric, which can affect the time scale, intensity and the pulse shape of the transient X-ray pulses. In the studies of `spot' case (Cheng, Li, \& Suen 1998; Tang \& Cheng 2001), Schwarzschild metric is used and all the effects of rotating metric are ignored.
For a complete analysis, development of an anisotropic transport equation is clearly needed and instead of Schwarzschild metric a rotating metric must be used.\\
\indent We organize the paper as follows. In section 2, we review the Hartle-Thorne formalism. In section 3, we study the effects of rotation on thermal equilibrium configuration. In section 4, we describe the development of the anisotropic heat transport equation. In section 5, we describe a Markovian random walk method to simulate
the thermal afterglows result from glitches. In section 6, we study the effects of the gravitational lensing and frame dragging on the X-ray light curves. In section 7, we discuss the physical interpretation of the numerical results.\\

\section{Hartle-Thorne Formalism}

The general expression for the line element of an axial symmetric spacetime is determined by time-translational invariance and
axial-rotational invariance:
\begin{equation}
ds^{2}=e^{2\nu \left( r,\theta \right) }dt^{2}-e^{2\lambda \left(
r,\theta \right) }dr^{2}-r^{2}e^{2\psi \left( r,\theta \right) }\left[
d\theta ^{2}+\sin ^{2}\theta \left( d\phi -\omega \left( r,\theta \right)
dt\right) ^{2}\right]
\end{equation}
We have chosen units $G=c=1$.\\
\indent To calculate the rotating stellar model as a perturbative expansion from a spherical star, we adopt the method proposed by Hartle (1967) and Hartle \& Thorne (1968). We calculate the perturbations up to second order of the rotational frequency. The perturbed geometry of spacetime is described by:
\begin{eqnarray}
ds^{2}&=&e^{2\nu \left(r\right)}\left[1+2\left(h_{0}+h_{2}P_{2}\left(\cos \theta\right)\right)\right]dt^{2}\nonumber\\
      &-&\frac{\left[1+2\left(2m_{0}+m_{2}P_{2}\left(\cos \theta\right)\right)/\left(r-2M\right)\right]}{1-2M/r}dr^{2}\nonumber\\
      &-&r^{2}\left[1+2\left(\nu_{2}-h_{2}\right)P_{2}\left(\cos \theta\right)\right]\left[d\theta^{2}+\sin^{2}\theta\left(d\phi-\omega dt\right)^{2}\right]\nonumber\\
      &+&O\left(\Omega^{3}\right)\nonumber\\
\end{eqnarray}
where $\Omega$ is the rotational frequency of the star, $P_{2}\left(\cos \theta\right)=(3\cos^{2} \theta-1)/2$ is the Legendre Polynomial of order 2 and $h_{0},h_{2},m_{0},m_{2},\nu_{2}$ are all
functions of $r$ that are proportional to $\Omega^{2}$ (Hartle 1967;  Hartle \& Thorne 1968). $\nu\left(r\right)$ and $M$ are the metric function and the gravitational mass of a non-rotating star respectively. With the chosen equation of state and the central density, they can be calculated by integrating the following set of equations
\begin{equation}
\frac{dP}{dr}=-\left(\rho+P\right)\frac{\left(M+4\pi r^{3}\right)}{r\left(r-2M\right)}
\end{equation}
\begin{equation}
\frac{dM}{dr}=4\pi r^{2}\rho
\end{equation}
\begin{equation}
\frac{d\nu}{dr}=-\frac{dP}{dr}\left(\rho+P\right)^{-1}
\end{equation}
outward from the center with the boundary conditions: $M\left(0\right)=0$; $P\left(R\right)=0$; $\nu\left(R\right)=\frac{1}{2}\ln\left(1-\frac{2M\left(R\right)}{R}\right)$, where $R$ is the radius of the non-rotating star.
For a uniformly and rigidly rotating star, the contravariant components of four velocity are:
\begin{equation}
u^{t}=e^{-\nu}\left[1+\frac{1}{2}r^{2}\sin^2\theta\bar{\omega}^{2}e^{-2\nu}-h_{0}-h_{2}P_{2}\left(\cos \theta\right)\right]
\end{equation}
\begin{equation}
u^{\phi}=\Omega u^{t}
\end{equation}
\begin{equation}
u^{r}=u^{\theta}=0
\end{equation}
where $\bar{\omega}=\Omega-\omega$ and $\omega$ defines the dragging frequency of local inertial frames.\\
\indent The star is centrifugally deformed by rotation. In the reference frame that is momentarily moving with the fluid, the energy-density
distribution including the effect of rotation is:
\begin{equation}
\rho+\Delta\rho=\rho+\left(\rho+P\right)\left(p_{0}^{*}+p_{2}^{*}P_{2}\left(\cos \theta\right)\right)\frac{d\rho}{dP}
\end{equation}
All the necessary rotational perturbation functions are calculated from the equations as described below.\\
\indent The angular velocity of the fluid relative to the local inertial frame $\bar{\omega}$, is found by integrating the differential equation:
\begin{equation}
\frac{1}{r^{4}}\frac{d}{dr}\left(r^{4}j\frac{d\bar{\omega}}{dr}\right)+\frac{4}{r}\frac{dj}{dr}\bar{\omega}=0
\end{equation}
where
\begin{equation}
j\left(r\right)=e^{-\nu}\left[1-\frac{2M}{r}\right]^{\frac{1}{2}}
\end{equation}
This equation can be integrated outward from the center of the star with the boundary conditions $\bar{\omega}=\bar{\omega}_{c}$ and
$\frac{d\bar{\omega}}{dr}=0$. The constant $\bar{\omega}_{c}$ is chosen arbitrarily. When the surface is reached, one can determine
the angular momentum $J=\frac{1}{6}R^{4}\left(\frac{d\bar{\omega}}{dr}\right)_{r=R}$ as well as the angular velocity $\Omega=\bar{\omega}\left(R\right)+\frac{2J}{R^{3}}$, corresponding to the chosen $\bar{\omega}_{c}$.\\
\indent $p_{0}^{*}$ and $m_{0}$ can be calculated by integrating the following equations
\begin{equation}
\frac{dm_{0}}{dr}=4\pi r^{2}\frac{d\rho}{dP}\left(\rho+P\right)p_{0}^{*}+\frac{1}{12}j^{2}r^{4}\left(\frac{d\bar{\omega}}{dr}\right)^{2}-\frac{1}{3}r^{3}\frac{dj^{2}}{dr}\bar{\omega}^{2}
\end{equation}
\begin{eqnarray}
\frac{dp_{0}^{*}}{dr}&=&-\frac{m_{0}\left(1+8\pi r^{2}P\right)}{\left(r-2M\right)^{2}}-\frac{4\pi\left(\rho+P\right)r^{2}}{\left(r-2M\right)}p_{0}^{*}\nonumber\\
                     &+&\frac{1}{12}\frac{r^{4}j^{2}}{\left(r-2M\right)}\left(\frac{d\bar{\omega}}{dr}\right)^{2}+\frac{1}{3}\frac{d}{dr}\left(\frac{r^{3}j^{2}\bar{\omega}^{2}}{r-2M}\right)\nonumber\\
\end{eqnarray}
outward from the center with boundary conditions $p_{0}^{*}=m_{0}=0$. Owing to the centrifugal force, a rotating star will have a larger mass than that of the non-rotating one for a given central density. The
increment is given as $\Delta M=m_{0}\left(R\right)+\frac{J^{2}}{R^{3}}$. $h_{0}$ can be obtained from the algebraic relations\\
Inside the star:
\begin{equation}
h_{0}=-p_{0}^{*}+\frac{1}{3}r^{2}e^{-2\nu}\bar{\omega}^{2}+h_{0c}
\end{equation}
Outside the star:
\begin{equation}
h_{0}=-\frac{\Delta M}{r-2M}+\frac{J^{2}}{r^{3}\left(r-2M\right)}
\end{equation}
$h_{0c}$ is a constant determined by demanding $h_{0}$ be continuous across the surface.\\
\indent The remaining four perturbation functions ($\nu_{2}, h_{2}, m_{2}, p_{2}^{*}$) can be calculated from the following equations
\begin{equation}
\frac{d\nu_{2}}{dr}=-2\frac{d\nu}{dr}h_{2}+\left(\frac{1}{r}+\frac{d\nu}{dr}\right)\left[-\frac{1}{3}r^{3}\frac{dj^{2}}{dr}\bar{\omega}^{2}+\frac{1}{6}j^{2}r^{4}\left(\frac{d\bar{\omega}}{dr}\right)^{2}\right]
\end{equation}
\begin{eqnarray}
\frac{dh_{2}}{dr}&=&\left[-2\frac{d\nu}{dr}+\frac{2r}{r-2M}\left(\frac{d\nu}{dr}\right)^{-1}\left[2\pi\left(\rho+P\right)-\frac{M}{r^{3}}\right]\right]h_{2}\nonumber\\
                 &-&\frac{2\nu_{2}}{r\left(r-2M\right)}\left(\frac{d\nu}{dr}\right)^{-1}\nonumber\\
                 &+&\frac{1}{6}\left[\frac{d\nu}{dr}r-\frac{1}{2\left(r-2M\right)}\left(\frac{d\nu}{dr}\right)^{-1}\right]r^{3}j^{2}\left(\frac{d\bar{\omega}}{dr}\right)^{2}\nonumber\\
                 &-&\frac{2}{3}\left[\frac{d\nu}{dr}r+\frac{1}{2\left(r-2M\right)}\left(\frac{d\nu}{dr}\right)^{-1}\right]r^{2}\bar{\omega}^{2}j\frac{dj}{dr}\nonumber\\
\end{eqnarray}
\begin{equation}
m_{2}=\left(r-2M\right)\left[-h_{2}-\frac{1}{3}r^{3}\left(\frac{dj^{2}}{dr}\right)\bar{\omega}^{2}+\frac{1}{6}r^{4}j^{2}\left(\frac{d\bar{\omega}}{dr}\right)^{2}\right]
\end{equation}
\begin{equation}
p_{2}^{*}=-h_{2}-\frac{1}{3}r^{2}e^{-\nu}\bar{\omega}
\end{equation}
with boundary conditions $\nu_{2}=0$ and $h_{2}=0$ at the origin.

\section{Thermal Equilibrium Configuration}

Consider a particle in the star with energy $E$, the energy of this particle as measured by a distant observer $E_{\infty}$ can be related to its
locally measured value $E$ by $E_{\infty}=\frac{E}{u^{t}}$. Next, we consider the energy transport from one region to another in an isolated star without any change in the density of particles.
For a local observer, the fundamental thermodynamics relation $dE=TdS$ is always valid, where $dS$ is the change in entropy per baryon and $T$ is the locally measured temperature.\\
\indent By following the scheme of Ze$\mbox{l}^{'}$dvoich and Novikov (1971), the equilibrium condition can be obtained by extremizing the entropy of the system with the constraint that the total energy is conserved:
\begin{eqnarray}
\delta\left(\int SndV+\Lambda\int E_{\infty}ndV\right)&=&0\nonumber\\
\int\left(\delta S+\Lambda\frac{\delta E}{u^{t}}\right)ndV&=&0\nonumber\\
\int\left(1+\Lambda\frac{T}{u^{t}}\right)ndV&=&0\nonumber\\
\end{eqnarray}
where $n$ is baryon number density, $dV$ is proper volume element and $\Lambda$ is a Lagrange multiplier. Consequently, the thermal equilibrium condition of a relativistic axisymmetric
star is formulated as:
\begin{equation}
\frac{T}{u^{t}}=\Lambda^{-1}=\mbox{constant}
\end{equation}
This result, which obtained from a relatively simple variation argument, is consistent with the one derived more vigorously by Miralles, Van Riper, \& Lattimer (1993).
Also, there is an underlying assumption that the neutron star is in rigid rotation. If it were not, the internal friction will produce heat and our variational argument is no longer valid.\\
\indent The aforementioned isothermal approximation is only appropriate in the interior. There is a thin atmosphere surrounds the isothermal core which
sustains an appreciable temperature gradient. For spherically symmetric cases, model atmospheres have already been calculated by Gudmundsson, Pethick, \& Epstein (1983), which give the surface temperature $T_{s}$ as a function of the
temperature at the base of the atmosphere $T_{b}$ and the surface gravity $g_{s}$. The base of the atmosphere is chosen at a particular density $\rho_{b}$. For our study, it is chosen at $\rho_{b}=10^{10}\mbox{ g cm}^{-3}$:
\begin{equation}
T_{b_{9}}=0.1288\left(\frac{T_{s_{6}}^{4}}{g_{s_{14}}}\right)^{0.455}
\end{equation}
\indent We shall assume this relation is valid in the slow rotation case with a suitable choice of an effective local surface gravity $g_{s}\left(\theta\right)$ that take the centrifugal deformation into account. Consequently, the surface
temperature $T_{s}\left(T_{b},\theta\right)$ depends on the polar angle.\\
\section{General Relativistic Anisotropic Heat Transport Equation}

With an intention to model the most general case of energy transport inside a rotating neutron star,
we derive the heat transport equation
without making the assumption of spherical symmetry in energy transport, stellar structure as well as the spacetime.\\
\indent To derive the equation, we assume that there are no other
entropy generating mechanisms besides diffusion and the fluid
motion inside the star due to the thermal effects is negligible.
The energy-momentum tensor inside a star consist of perfect fluid
which allow heat flow can be written as
$T^{\mu\nu}=\left(
\rho+P\right)
u^{\mu}u^{\nu}-Pg^{\mu\nu}+u^{\mu}q^{\nu}+u^{\nu}q^{\mu}$, where
$g^{\mu\nu}$ denotes the inverse components of equation (1),
$u^{\mu}$ is the 4-velocity of the fluid flow which has been
described in section 2, $\rho$ and $P$ are the total energy
density and pressure measured in the rest frame of the fluid{\bf
(Tolman 1934)}. The heat flow is given by
$q^{\alpha}=K\left(g^{\alpha\beta}-u^{\alpha}u^{\beta}\right)\left(T_{;\beta}-Ta_{\beta}\right)$,
where $a_{\alpha}=u_{\alpha;\beta}u^{\beta}$ is the 4-acceleration and $K$ is the thermal conductivity{\bf (Tolman 1934)}.\\
\indent No matter how complicated the transport equation is, the basic principle is that the underlying physics is
nothing but the law of
conservation of energy. In relativity, the energy-momentum tensor embodies a compact description of energy
and momentum. To be more
specific, the transport equation can be derived by the conservation of energy-momentum tensor, namely,
\begin{equation}
T^{t\alpha}_{;\alpha}=T^{tt}_{;t}+T^{tr}_{;r}+T^{t\theta}_{;\theta}+T^{t\phi}_{;\phi}=0
\end{equation}
where
\begin{eqnarray}
T_{;t}^{tt}&=&\left[u^{t}u^{t}+\left(u^{t}u^{t}-g^{tt}\right)\frac{dP}{d\rho}\right]
\frac{\partial \rho}{\partial t}\nonumber\\
&+&2u^{t}K\left[\left(g^{tt}-u^{t}u^{t}\right)\frac{\partial^{2}T}{\partial t^{2}}+\left(g^{t\phi}-u^{t}u^{\phi}\right)\frac{\partial^{2}T}{\partial t\partial \phi}\right]\nonumber\\
&+&2\Gamma_{rt}^{t}f^{r}+2\Gamma_{\theta t}^{t}f^{\theta}
\end{eqnarray}
\begin{equation}
T_{;r}^{tr}=\frac{\partial f^{r}}{\partial r}+\frac{\partial\lambda}{\partial r}f^{r}
+\frac{\partial\lambda}{\partial\theta}f^{\theta}
+\Gamma_{tr}^{t}f^{r}+\Gamma_{r\phi}^{t}u^{\phi}q^{r}\nonumber\\
\end{equation}
\begin{equation}
T_{;\theta}^{t\theta}=\frac{\partial f^{\theta}}{\partial\theta}+\Gamma_{t\theta}^{t}f^{\theta}+\Gamma_{r\theta}^{\theta}f^{r}
+\Gamma_{\theta\theta}^{\theta}f^{\theta}+\Gamma_{\theta\phi}^{t}u^{\phi}q^{\theta}\nonumber\\
\end{equation}
\begin{equation}
T_{;\phi}^{t\phi}=\frac{\partial f^{\phi}}{\partial\phi}+\Gamma_{\theta\phi}^{\phi}f^{\theta}+\Gamma_{r\phi}^{\phi}f^{r}
+\Gamma_{\theta\phi}^{t}u^{\phi}q^{\theta}+\Gamma_{r\phi}^{t}u^{\phi}q^{r}
\end{equation}
where $f^{i}=T^{ti}$ is the energy flow per unit area parallel to $i$ direction. $\Gamma _{\alpha\beta}^{\gamma}$ are the Christoffel symbols. $\frac{\partial \rho}{\partial t}$ in $T_{;t}^{tt}$ is the rate of change of energy density measured by
distant observer, which can be expressed in the locally measured quantity, namely, $\frac{\partial \rho}{\partial \tau}$. It depends on the processes under consideration.
If only heat conduction and neutrino emission are considered, we have $\frac{\partial\rho}{\partial\tau}=C_{v}\frac{\partial T}{\partial\tau}+Q_{\nu}$,
where $C_{v}$ and $Q_{\nu}$ are the heat capacity and the neutrino emissivity measured in the local frame respectively. Expressing it in the coordinate time $t$, we have
$\frac{\partial\rho}{\partial t}=C_{v}\frac{\partial T}{\partial t}+\frac{Q_{\nu}}{u^{t}}$.
Using this relation and equation (23), we obtain the heat diffusion equation.
\begin{eqnarray}
0&=&\left[u^{t}u^{t}+\left(u^{t}u^{t}-g^{tt}\right)\frac{dP}{d\rho}\right]
\left(C_{v}\frac{\partial T}{\partial t}+\frac{Q_{\nu}}{u^{t}}\right)\nonumber\\
&+&2u^{t}K\left[\left(g^{tt}-u^{t}u^{t}\right)\frac{\partial^{2}T}{\partial t^{2}}
+\left(g^{t\phi}-u^{t}u^{\phi}\right)\frac{\partial^{2}T}{\partial t\partial \phi}\right]\nonumber\\
&+&\frac{\partial f^{r}}{\partial r}+\frac{\partial f^{\theta}}{\partial\theta}+\frac{\partial f^{\phi}}{\partial\phi}\nonumber\\
&+&\left(3\Gamma_{rt}^{t}+\frac{\partial\lambda}{\partial r}+\Gamma_{r\theta}^{\theta}+\Gamma_{r\phi}^{\phi}\right)f^{r}\nonumber\\
&+&\left(3\Gamma_{t\theta}^{t}+\frac{\partial\lambda}{\partial\theta}+\Gamma_{\theta\theta}^{\theta}+\Gamma_{\theta\phi}^{\phi}\right)f^{\theta}\nonumber\\
&+&2\Gamma_{r\phi}^{t}u^{\phi}q^{r}+2\Gamma_{\theta\phi}^{t}u^{\phi}q^{\theta}
\end{eqnarray}
\indent In the non-rotational limit, the spacetime is described by the diagonal Schwarzschild metric, the above equation recovers the one obtained by Cheng, Li, \& Suen (1998).
It should also be noted that this equation recovers the well-known Newtonian case.

\section{Simulations of Thermal Afterglows result from Glitches}

{\bf The anisotropic relativistic heat transport equation, i.e. equation (28), can be rearranged into a single variable partial differential equation (cf. Appendix).} We employ a quantized Monte Carlo technique for the simulation of heat transport inside a rotating neutron star. Dynamic stochastic processes are simulated by using rate coefficients (i.e.
diffusion coefficient $D$, drift coefficient $v$, annihilation coefficient $\mu$ and local depletion $S$). These coefficients are determined via rearranging the rate equation (28).
In order to perform the simulations, some assumptions must be made. First, we assume that the quantities $D_{i}, v_{i} \mbox{ and } \mu_{i}$
depend only on position where $i$ denotes the motion of energy carriers in $r, \theta \mbox{ and } \phi$. Much more general dependences result in non-linear behavior.
Second, if the grid sizes are fine enough, the energy carriers transit from the surrounding grid points in the previous time step with equal probability.
Third, we neglect the thermal effects on the stellar structure.\\
\indent There are two terms of mixed derivative and second time derivative present in the equation. There is no easy way to incorporate them in a random walk approach. However, their coefficients only contain the factors of $(g^{tt}-u^{t}u^{t})$ and $(g^{t\phi}-u^{t}u^{\phi})$. Since the zero order terms are cancelled in the subtractions,
only second order perturbation terms are remained. Up to the highest rotational frequency we have considered ($1\times 10^{3} s^{-1}$), these coefficients are still negligible in comparison with the smallest term we have taken into account. So it is safe to drop it in the simulations.\\
\indent With these assumptions, the heat transport inside a rotating neutron star can now be simulated by a straightforward Markovian random walk (Mackeown 1997):
(1) Specify the initial condition $T\left(\vec{r},0\right)$ and quantize it by a large number, $M$, of ``walkers" whose starting positions are selected in accordance with $T\left(\vec{r},0\right)$ and their random walk is modeled.
(2) In each time step $\Delta t$, a weight is attached to each ``walker" according to the survival probability $\mu\Delta t$ by means of \emph{survival biasing} (Mackeown 1997). Then, it is displaced by $\Delta x_{i}=\pm\sqrt{2D_{i}\Delta t}+v_{i}\Delta t$
in each degree of freedom, where the sign of the step is chosen with equal probability on the basis of a uniform random variate.
(3) Incorporate the local depletion $S\Delta t$. (4) Repeat steps 2 and 3 $M$ times for each ``walker". (5) Repeat step 4 for N times.
(6) Plot the distribution of the ``walkers" and this should approximate the solution $T\left(\vec{r},t\right)$ at $t=N\Delta t$.
Note the finite value of ``walkers" $M$ introduces an unavoidable statistical fluctuation. Also, the finite time step used will introduce a truncation error.\\
\indent Since we intent to investigate the effect of rotation on the heat transport, we adopt a simple model of neutron star. We employ the method described in section 2 to calculate the rotational
stellar structure. The stellar structure is determined by the equation of state. We use the equation of state of neutron matter from Pandharipande (1971) in our calculation. We have considered the contribution of protons, neutrons and electrons to the heat capacity (Maxwell 1979).
But the most important term is the contribution of electrons. We only consider the neutrino emissivities due to neutron-neutron bremsstrahlung, proton-neutron bremsstrahlung and the modified URCA process (Maxwell 1979). We adopt the analytic formulae of
thermal conductivity presented by Flower \& Itoh (1981).\\
\indent We calculate temperature distributions for `ring' case and `spot' case. For both cases, heat inputs are deposited at the depth of the crust where $\rho \sim 10^{13}$ g $\mbox{cm}^{-3}$. The ratio $R/M\sim 8$ corresponds to $M\sim 1 M_{\odot}$ and $R\sim 10$ km.
The core temperature is taken to be $10^{7}$K. With $\Delta E=10^{42}\mbox{ergs}$ deposited around the equator (i.e. $\theta=90^{\circ}$), figure 1 illustrates the polar angular surface temperature distribution of a rotational `ring' case with $\Omega=7.6\times 10^{2}\mbox{s}^{-1}$.
For the `spot' case, we choose $\Delta E=10^{42}\mbox{ergs}$ to be released at $\theta=\phi=90^{\circ}$. Figure 2 illustrates the azimuthal profile of surface temperature at the rotational equator of the hot spot for a rotational case with $\Omega=7.6\times 10^{2}\mbox{s}^{-1}$.
Figure 3 illustrates the evolution of thermal X-ray flux of `spot' case. Three cases with different rotational frequency are compared.
With the mass increased by $\sim 33\%$ and radius decreased by $\sim 1\%$, the $R/M$ ratio is reduced to $\sim 6$ which indicates the gravitational effect is enhanced. We re-calculate the `spot' case with this $R/M$ ratio and the results for three cases with different rotational frequency are shown in figure 4.
The time scale and the magnitude of the afterglows for the non-rotational cases are consistent with those obtained by pervious authors (Tang \& Cheng 2001).\\
\indent The heat propagation time $\tau$ is approximately inversely proportional
to the diffusion coefficient $D_{r}$, which is proportional to thermal conductivity $K$ and inversely proportional to heat capacity $C_{v}$. In the region we are interested, the heat
capacity is proportional to temperature, while the thermal conductivity is inversely proportional to the temperature. Hence, $\tau$ is roughly proportional to the temperature squared. It can
be seen that heat propagation becomes much slower as the temperature becomes higher. For this reason, the duration of afterglow for the `ring' case is shorter than that for the `spot' case for the same heat input.
Comparing the non-rotational flux curve of `spot' case with $R/M\sim 8$ in figure 3 and that with $R/M\sim 6$ in figure 4, we find that the afterglow take a longer time in the model with $R/M\sim 6$
because a larger mass underneath the location of energy release that hinder the propagation of heat pulse to the surface. Comparing the rotational cases with the non-rotational one in figure 3 \& 4, we find that the duration of thermal afterglow is shortened when the effect of rotation is introduced.
To further investigate this effect, we carry out more simulations of `spot' case with different rotational frequency.
Figure 5 illustrates the fractional decrease in the duration of thermal afterglow as a function of rotational frequency for $R/M\sim 8$ and $R/M\sim 6$.\\
\indent The rotational effects can be originated from the centrifugal forces in rotating neutron stars. When the energy $\Delta E$ is released in a localized region at the equator, it can be viewed as an equivalent mass $\sim\Delta E/c^{2}$ that corotates with the star. This leads to a natural interpretation that centrifugal forces prompt a faster and larger thermal response. In figure 5,
it should be noticed that the fractional decrease in the duration is larger in the case of $R/M\sim 6$ than that in the case of $R/M\sim 8$ for a given $\Omega$. This indicates the other effect of rotation on the heat transfer. Apart from prompting a faster heat propagation, centrifugal forces also deform the star.
As pointed out by some ellipticity studies (Abramowicz 1990; Chandrasekhar \& Miller 1974), the denser a neutron star, the less eccentric it is. The star with $R/M\sim 8$ is more eccentric than the one with $R/M\sim 6$.
Since the heat inputs are deposited at the depth where $\rho\sim 10^{13}\mbox{g cm}^{-3}$ at the equator in both models, the heat pulse needs to propagate through a larger distance in a more eccentric star in order to reach the surface and give the afterglow.\\
\indent Nevertheless, the behavior of centrifugal forces in general relativity is fundamentally different from that in Newtonian physics (Freire \& Costa 1999; Abramowicz, Carter, \& Lasota 1988; Abramowicz \& Prasanna 1990; Abramowicz, Miller, \& Stuchl\'{i}k 1993). One of the questions behind our interpretation is whether there can be inversion of centrifugal forces exist inside neutron stars.
Freire \& Costa (1999) show that most realistic equations of state do not allow the existence of such inversion. Without the bother of centrifugal force inversion, we suggest that rotation introduces two effects on the thermal afterglow. First, it prompts a faster and larger thermal response. Second, it increases the distance that heat need to travel through in order to give an afterglow on the surface. The resultant effect
on the duration and the magnitude of the thermal afterglow depends on the interplay between these two factors.

\section{Thermal X-ray Light Curves}
For the `spot' case, the locally released energy would modulate the X-ray pulse shape by heating a portion of crust so that more thermal X-rays are emitted at a particular phase. The modulation of X-ray pulse will last until the surface temperature equilibrated. We have investigated the effects of spacetime curvature and rotation on the thermal X-ray profile.
\subsection{Gravitational Lensing Effect}
Since the surface gravity of a neutron star is tremendous, the effect of gravity on the trajectory of emitted photons must be taken into account. In this section, we choose our coordinates so that the observer is on the positive $z$-axis at $r=r_{0}$ where $r_{0}\rightarrow\infty$ (see figure 6).
The following scheme of calculating the X-ray light curves resulting from light bending is adapted from Pechenick et al. (1983).
The surface of the star is described by angular spherical coordinates, $\theta$ and $\phi$, where $\theta$ is measured from the $z$-axis we have just defined. When the photon is being emitted from the surface at an angle $\delta$ as illustrated in figure 6, it will be deflected by the gravitational field.
It will seem to the observer that it is emitted at angle $\theta^{'}$ from the $z$-axis. Hence, $\theta$ is a function of $\theta^{'}$:
\begin{equation}
\theta=\int_0^\frac{M}{R}\left[\left(\frac{M}{b}\right)^{2}-\left(1-2u\right)u^{2}\right]^{-\frac{1}{2}}\,du
\end{equation}
where $b=r_{0}\theta^{'}$ and $u=M/r$. $b$ is the impact parameter of the photon.\\
\indent If the star has the ratio $R/M>3$, then a photon emitted from the surface and reach the observer must have an impact parameter $b\leq b_{max}$, where $b_{max}=R\left(1-\frac{2M}{R}\right)^{-\frac{1}{2}}$
The maximum possible value of $\theta$ occurs when $b=b_{max}$.
We consider a hot spot of angular radius $\alpha$ centered at $\theta = \theta_{0}$. A function $h(\theta ; \alpha ,\theta_{0})$ is then defined as the range of $\phi$ included in the ``one-dimensional slice" at $\theta$ of the hot spot.
If $\theta_{0}+\alpha\leq\theta_{max}\leq 180^{o}$ and $\theta_{0}-\alpha\geq 0$, then $h(\theta ; \alpha ,\theta_{0})$ is defined as:
\begin{equation}
h(\theta ; \alpha ,\theta_{0})=\left\{\begin{array}{ll}
                               2\cos^{-1}\left(\frac{\cos\alpha -\cos\theta_{0}\cos\theta}{\sin\theta_{0}\sin\theta}\right) & \mbox{for $\theta_{0}-\alpha\leq\theta\leq\theta_{0}+\alpha$};\\
                               0 & \mbox{for $\theta$ outside the range $\theta_{0}\pm\alpha$}\end{array}\right.
\end{equation}
When the requirement that the photons under consideration must reach the observer is imposed (not all the photons emitted at $(\theta ,\phi)$ can reach the observer), $\delta$ is also a function of $\theta^{'}$. Let $x=\frac{b}{M}$ and $x_{max}=\frac{b_{max}}{M}$, we have
\begin{equation}
\delta=\sin^{-1}\left(\frac{x}{x_{max}}\right)
\end{equation}
We will come back to this requirement when we discuss the rotational effect on the pulse shape.\\
\indent In order to obtain light curves, $\theta_{0}$ has to be obtained as a function of time. If $\beta$ is the angle between the axis of rotation of the star and the line joining the center of the hot spot and the center of the star, and $\gamma$ is the angle between the axis of rotation and the $z$-axis (see figure 7), then
\begin{equation}
\cos\theta_{0}=\sin\left(\beta\right)\sin\left(\gamma\right)\cos\left(\Omega t\right) + \cos\left(\beta\right)\cos\left(\gamma\right)
\end{equation}
where $\Omega$ is the rotational frequency of the star.
We are now able to calculate the relative brightness:
\begin{equation}
A\left(\theta_{0};f,M/R,\alpha\right)=\left(1-\frac{2M}{R}\right)^{2}\left(\frac{M}{R}\right)^{2}\int_0^{x_{max}} f\left(\delta\left(x\right)\right)h\left(x; \alpha ,\theta_{0}\right)x\,dx
\end{equation}
where $f\left(\delta\right)=1$ for isotropic emission, $f\left(\delta\right)=\cos\delta$ for enhanced emission and $f\left(\delta\right)=\sin\delta$ for suppressed emission. The relationship between $A$ and $\Omega t$ is plotted in figure 8.\\

\subsection{Rotational Effect}
Apart from getting bent due to spacetime curvature, the trajectory of a photon emitted in a general direction from a point $r_{e}$ will be dragged away from its original direction of emission when the rotation of the relativistic star is taken into consideration (Kapoor \& Datta 1985).
In this section, the axis of rotation is taken to be $z$-axis. The net angle of deflection in the trajectory will be given by:
\begin{equation}
-\phi_{0}=\int_{r_{e}}^{r_{0}}\frac{\omega\left(1+\omega q_{e}\right)-q_{e}e^{2\nu}-e^{2\psi}}{e^{\nu -\lambda}\left[\left(1+\omega q_{e}\right)^{2}-q_{e}^{2}e^{2\nu}-e^{2\psi}\right]^{\frac{1}{2}}}\,dr
\end{equation}
with $r_{e}$ and $r_{0}$ denote the point of emission and the observer's location respectively. $e^{\nu}$, $e^{\psi}$ and $e^{\lambda}$ are the exterior rotating metric components:
\begin{equation}
e^{2\nu}=e^{-2\lambda}=1-\frac{2M}{r}+\frac{2J^{2}}{r^{4}}
\end{equation}
\begin{equation}
e^{2\psi}=r^{2}\sin^{2}\theta
\end{equation}
The dragging frequency $\omega$ has an analytic exterior solution:
\begin{equation}
\omega\left(r\right)=\frac{2J}{r^{3}}
\end{equation}
$q_{e}$ in the integrand of equation (34) is defined as the impact parameter of the photon with rotation taken into consideration:
\begin{equation}
q_{e}=\left.\frac{e^{\psi -\nu}\left(e^{\psi -\nu}\left(\Omega-\omega\right)+\sin\delta\right)}{1+e^{\psi -\nu}\left(e^{\psi -\nu}\omega\left(\Omega-\omega\right)+\Omega\sin\delta\right)}\right|_{r=r_{e}}
\end{equation}
In our case, the source (hot spot) is located at the rotational equator, $\delta$ represents the azimuthal angle at which the photon is emitted with respect to the normal vector of the surface as seen in the local rest frame of the star. We choose the convention that $\delta=0$ for a radial outgoing photon,
$\delta <0$ for a tangentially forward photon, and $\delta >0$ for a tangentially backward photon.\\
\indent As mentioned in the section 6.2, only the photons emitted at a particular value of $\delta$ would be received by the distant observer. The value of $q_{e}$ has been obtained by imposing this requirement. Because of the effect of frame dragging around the neutron star, the situation will be different from that
in previous section.\\
\indent It should be noted that the impact parameter $q_{e}$ is not symmetric in $\pm\delta$. In principle, this asymmetry will manifest itself in the final pulse shape:
\begin{equation}
\delta_{new}=\delta +\phi_{0}\left(\delta\right)
\end{equation}
Figure 9 schematically illustrates the deflection of a photon trajectory. As a consequence of rotation, $\left|\phi_{0}\left(\delta\right)\right|\not=\left|\phi_{0}\left(-\delta\right)\right|$.\\
\indent This asymmetry does not occur in spherically symmetric spacetime, where although the spacetime curvature will still deflect the trajectory of photon. This can be easily seen if one approximates the corresponding Schwarzschild expression of equation (34) as follows:
\begin{equation}
\phi^{S}_{0}\approx -\frac{q^{S}}{R}\left[1+\frac{\sin ^{2}\delta\left(2R-3R^{S}\right)}{12\left(R-2R^{S}\right)}\right]
\end{equation}
where the superscript $S$ refers to the Schwarzschild case. Here $\theta=90^{o}$ for the reason of symmetry. $R^{S}$ denotes Schwarzschild radius and $q^{S}=\left(e^{\psi-\nu}\sin\delta\right)_{r=R}$.
It can be seen that $\left|\phi^{S}_{0}\left(\delta\right)\right|=\left|\phi^{S}_{0}\left(-\delta\right)\right|$.\\
\indent We define the corresponding $\delta^{S}_{new}$ of equation (39) as
\begin{equation}
\delta^{S}_{new}=\delta +\phi^{S}_{0}\left(\delta\right)
\end{equation}
comparison between $\delta^{S}_{new}$ and $\delta_{new}$ is made in figure 10. Even in the fastest rotational case we have considered ($\Omega=1\times 10^{3}\mbox{s}^{-1}$) with a relatively small gravitational effect ($R/M\sim 8$), the comparison indicates that the effect of rotation is small and the effect of spacetime curvature predominates.\\

\subsection{X-ray Light Curves}
\indent\\
\indent We can generate the light curves in soft X-ray regime for different `spot' cases. According to Pechenick et. al. (1983), the total energy flux observed is:
\begin{equation}
F_{X}=\sum I_{0}\left(\frac{R}{r_{0}}\right)^{2}A\left(\theta_{0};f,M/R,\alpha\right)
\end{equation}
where $I_{0}$ is the energy flux from each cell at the surface including the factor of enhanced emission. We adopt $\delta_{new}$ instead of $\delta^{S}_{new}$ to incorporate the asymmetry due to rotation. $\sum$ denoted the summation of the contribution from each cell.
The comparisons of light curves with different rotational frequency are made in figure 11 and figure 12 with $R/M\sim 8$ and $R/M\sim 6$ respectively. We have not found any asymmetry in the pulse profile for the rotational cases.\\

\section{Conclusion and Discussion}
In this paper, we have developed an anisotropic heat transport equation. This equation is then employed to simulate thermal afterglows resulting from pulsar glitches. We examine the effects of rotation on the thermal responses. Although rotation does not have significant effect on the rate of
standard cooling (Miralles, Van Riper, \& Lattimer 1993), noticeable changes on duration and intensity of thermal afterglows are found in `spot' cases. In comparison with static cases, afterglows with shorter duration and larger intensity are found in rotating stars. These are not unexpected when the effects of centrifugal forces on the stellar
structure as well as heat transport are realized. We suggest that rotation prompts a faster and larger response. {\bf In equation A1(cf. Appendix), the radial diffusion coefficient is given by $D_{r}=-\frac{C_{2}}{C_{6}}$ which contains the factors of rotating metric and hence is a function of rotational frequency. $D_{r}$ increases with rotational frequency (see table 1) and prompts a faster response.}  On the other hand, the rotation also increases the distance that heat pulses need to travel through in order to give an afterglow on the surface. The resultant effect on the duration and the magnitude of the thermal afterglow depends on the interplay between these two factors.
We have also generated the thermal X-ray light curves for the `spot' cases. The effects of gravitational lensing and frame dragging are fully taken into account. Apart from the peak time and the peak intensity, we have not found any significant effect of rotation on the morphology of the pulse profile.\\
\indent Hirano et al. (1997) characterize the thermal afterglow by the fractional increase of surface temperature at the peak and the peak time. They also propose a framework to set constraint on the equation of state by using these two parameters. Since the peak time is larger and the amplitude is smaller for a stiff star than a soft star. Without taking the effects of rotation into consideration,
we may underestimate the stiffness of the equation of state.\\
\indent One of the practical questions is the detectability of these soft X-ray transients which depends on
the sensitivity of the detector. For the `spot' cases considered in this paper, the glitch events are able
to be detected by the state of the art X-ray satellites such as XMM-Newton. Once the observational data of
these thermal afterglows are obtained, these can
be used in putting constraints on the equation of state of neutron stars as well as the glitch models by
the method of periodic analysis (Andersen \& \"{O}gelman 1997).\\
\indent In our calculations, we have neglected interstellar
absorption, magnetospheric effects, magnetic field effect (Page
1995) and the possibility of uplifting and local expansion of
matter caused by the heat deposition (Eichler \& Cheng 1989). {\bf
We would like to remark that we have ignored the contribution of impurity
scattering in the thermal conductivity, which may be important when $T \leq 10^7$ (Yakovlev \& Urpin 1980).} We have also assumed that
there is no other glitch events occur during the afterglow is
evolving. If the glitch events occur too frequently, it may result
in a pile-up of pulses as well as a long term variation of the
total thermal radiation which eventually reduces the detectability
of the thermal afterglow (Li 1997). Moreover, we keep a constant
rotational frequency during the afterglow is evolving. However,
glitch recovers exponentially and hence $\Omega$ should varies
with time. Also, the effects should be more important for fast
rotation, a fully numerical scheme is needed to calculate the
metric in fast rotation case. For further study, all these effects
have to be taken into account.

\appendix
\section{Appendix}
\indent\\
\indent The anisotropic relativistic heat transport equation (i.e. equation (28)) can be rearranged into a standard form of partial differential equation as follows:
\begin{equation}
0=C_{1}\frac{\partial^{2}T}{\partial t^{2}}+C_{2}\frac{\partial^{2}T}{\partial r^{2}}+C_{3}\frac{\partial^{2}T}{\partial \theta^{2}}+C_{4}\frac{\partial^{2}T}{\partial \phi^{2}}+
C_{5}\frac{\partial^{2}T}{\partial t\partial\phi}+C_{6}\frac{\partial T}{\partial t}+C_{7}\frac{\partial T}{\partial r}+C_{8}\frac{\partial T}{\partial\theta}+C_{9}\frac{\partial T}{\partial\phi}+
C_{10}T+C_{11}
\end{equation}
The expressions of coefficients $C_{1}$ to $C_{11}$ are listed below (with $G=c=1$):
\begin{equation}
C_{1}=2u^{t}K\left(g^{tt}-u^{t}u^{t}\right)
\end{equation}
\begin{equation}
C_{2}=u^{t}Kg^{rr}
\end{equation}
\begin{equation}
C_{3}=u^{t}Kg^{\theta\theta}
\end{equation}
\begin{equation}
C_{4}=K\left[u^{t}\left(g^{\phi\phi}-u^{\phi}u^{\phi}\right)+u^{\phi}\left(g^{t\phi}-u^{t}u^{\phi}\right)\right]
\end{equation}
\begin{equation}
C_{5}=K\left[3u^{t}\left(g^{t\phi}-u^{\phi}u^{t}\right)+u^{\phi}\left(g^{tt}-u^{t}u^{t}\right)\right]
\end{equation}
\begin{equation}
C_{6}=\frac{\partial K}{\partial\phi}\left[u^{t}\left(g^{t\phi}-u^{\phi}u^{t}\right)+u^{\phi}\left(g^{tt}-u^{t}u^{t}\right)\right]+
C_{v}\left[\frac{dP}{d\rho}\left(u^{t}u^{t}-g^{tt}\right)+u^{t}u^{t}\right]
\end{equation}
\begin{equation}
C_{7}=K\frac{\partial u^{t}}{\partial r}g^{rr}+u^{t}\frac{\partial K}{\partial r}g^{rr}+u^{t}K\frac{\partial g^{rr}}{\partial r}+
\left(3\Gamma^{t}_{rt}+\frac{\partial\lambda}{\partial r}+\Gamma^{\theta}_{r\theta}+\Gamma^{\phi}_{r\phi}\right)u^{t}Kg^{rr}
+2\Gamma^{t}_{r\phi}u^{\phi}Kg^{rr}-u^{t}Kg^{rr}a_{r}
\end{equation}
\begin{equation}
C_{8}=K\frac{\partial u^{t}}{\partial\theta}g^{\theta\theta}+u^{t}\frac{\partial K}{\partial\theta}g^{\theta\theta}+u^{t}K\frac{\partial g^{\theta\theta}}{\partial\theta}+
\left(3\Gamma^{t}_{\theta t}+\frac{\partial\lambda}{\partial\theta}+\Gamma^{\theta}_{\theta\theta}+\Gamma^{\phi}_{\theta\phi}\right)u^{t}Kg^{\theta\theta}
-u^{t}Kg^{\theta\theta}a_{\theta}
\end{equation}
\begin{equation}
C_{9}=\frac{\partial K}{\partial\phi}\left[u^{t}\left(g^{\phi\phi}-u^{\phi}u^{\phi}\right)+u^{\phi}\left(g^{t\phi}-u^{t}u^{\phi}\right)\right]
\end{equation}
\begin{equation}
C_{10}=-\left[a_{r}\left(C_{7}+u^{t}Kg^{rr}a_{r}\right)+a_{\theta}\left(C_{8}+u^{t}Kg^{\theta\theta}a_{\theta}\right)+u^{t}Kg^{rr}\frac{\partial a_{r}}{\partial r}+u^{t}Kg^{\theta\theta}\frac{\partial a_{\theta}}{\partial\theta}\right]
\end{equation}
\begin{equation}
C_{11}=\frac{Q_{\nu}}{u^{t}}\left[\frac{dP}{d\rho}\left(u^{t}u^{t}-g^{tt}\right)+u^{t}u^{t}\right]
\end{equation}

\clearpage
\begin{figure}
\plotone{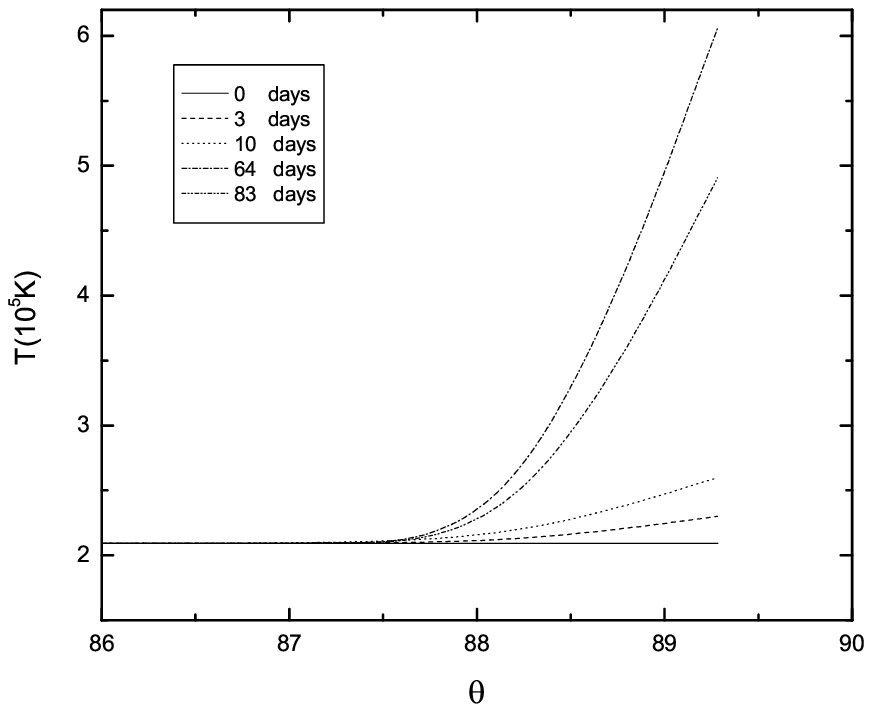}
\caption{Surface temperature of a `ring' case as a function of polar angle for a neutron star with $R/M\sim 8$, $\Omega=7.6\times 10^{2}\mbox{s}^{-1}$, $T_{c}=10^{7} \mbox{K}$, $\Delta E=10^{42}\mbox{erg}$ at $\rho_{glitch}=10^{13}\mbox{g cm}^{-3}$, $\theta=90^{\circ}$. The upper panel denotes days elapsed after the heat deposition.}
\end{figure}

\clearpage
\begin{figure}
\plotone{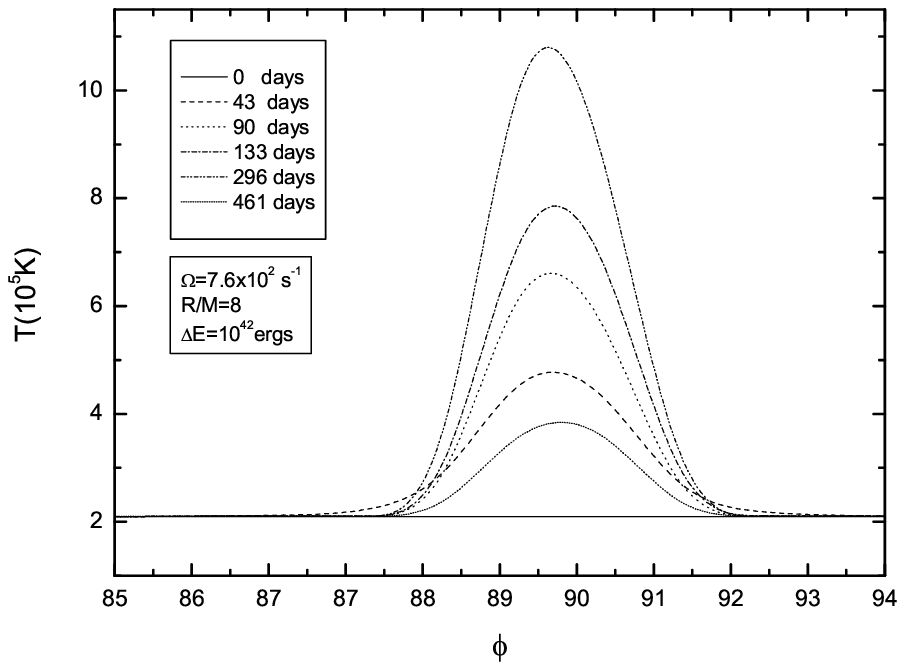}
\caption{Surface temperature of a `spot' case at the equator as a function of azimuthal angle for a neutron star with $R/M\sim 8$, $\Omega=7.6\times 10^{2}\mbox{s}^{-1}$, $T_{c}=10^{7} \mbox{K}$, $\Delta E=10^{42}\mbox{erg}$ at $\rho_{glitch}=10^{13}\mbox{g cm}^{-3}$, $\theta=\phi=90^{\circ}$. The upper panel denotes days elapsed after the heat deposition}
\end{figure}

\clearpage
\begin{figure}
\plotone{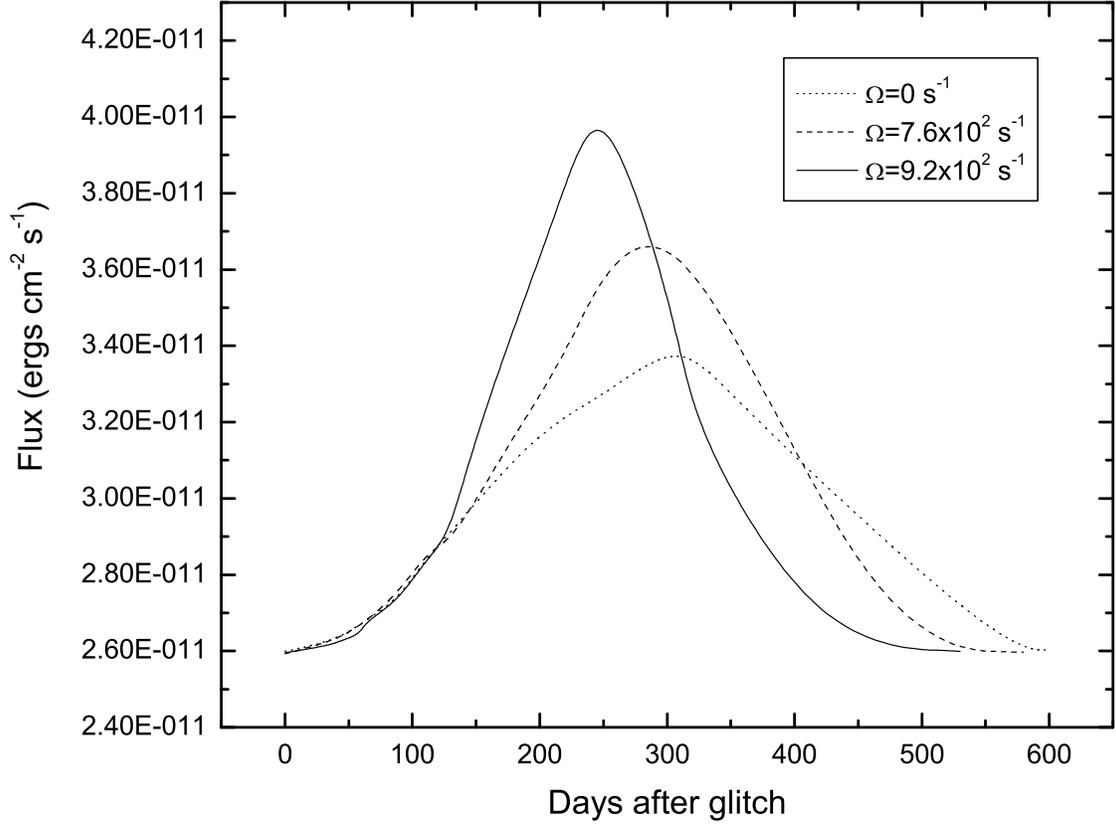}
\caption{The evolution curves of thermal X-ray flux for a neutron star with $R/M\sim 8$, $T_{c}=10^{7} \mbox{K}$, $\Delta E=10^{42}\mbox{erg}$ at $\rho_{glitch}=10^{13}\mbox{g cm}^{-3}$, $\theta=\phi=90^{\circ}$ for a `spot' case. Three cases with different rotational frequency are compared. }
\end{figure}

\clearpage
\begin{figure}
\plotone{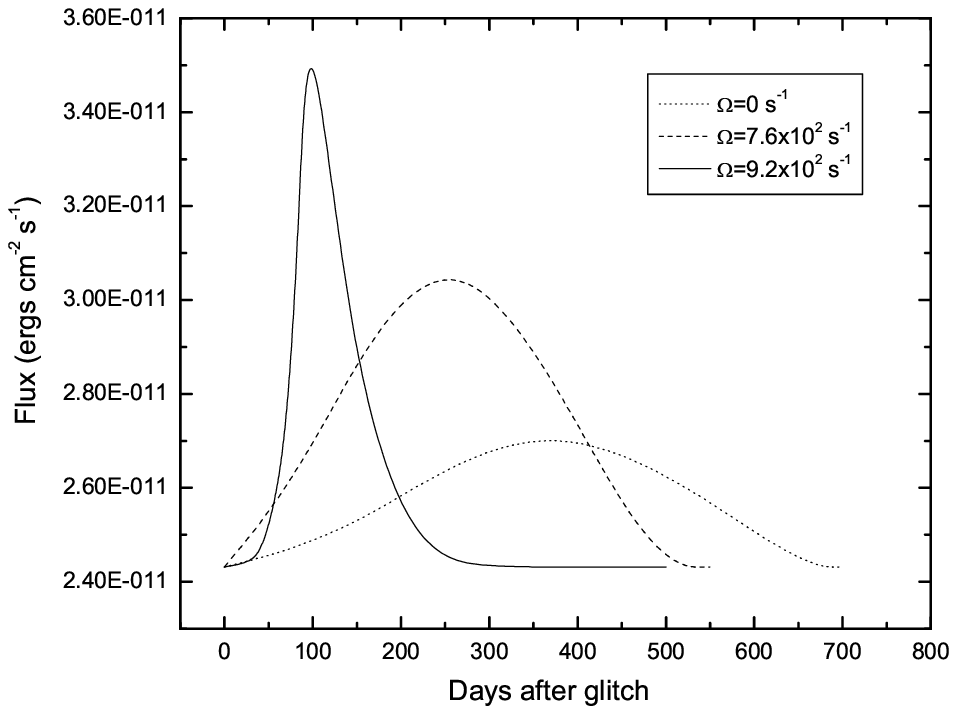}
\caption{Same as figure 3 but with $R/M\sim 6$. }
\end{figure}

\clearpage
\begin{figure}
\plotone{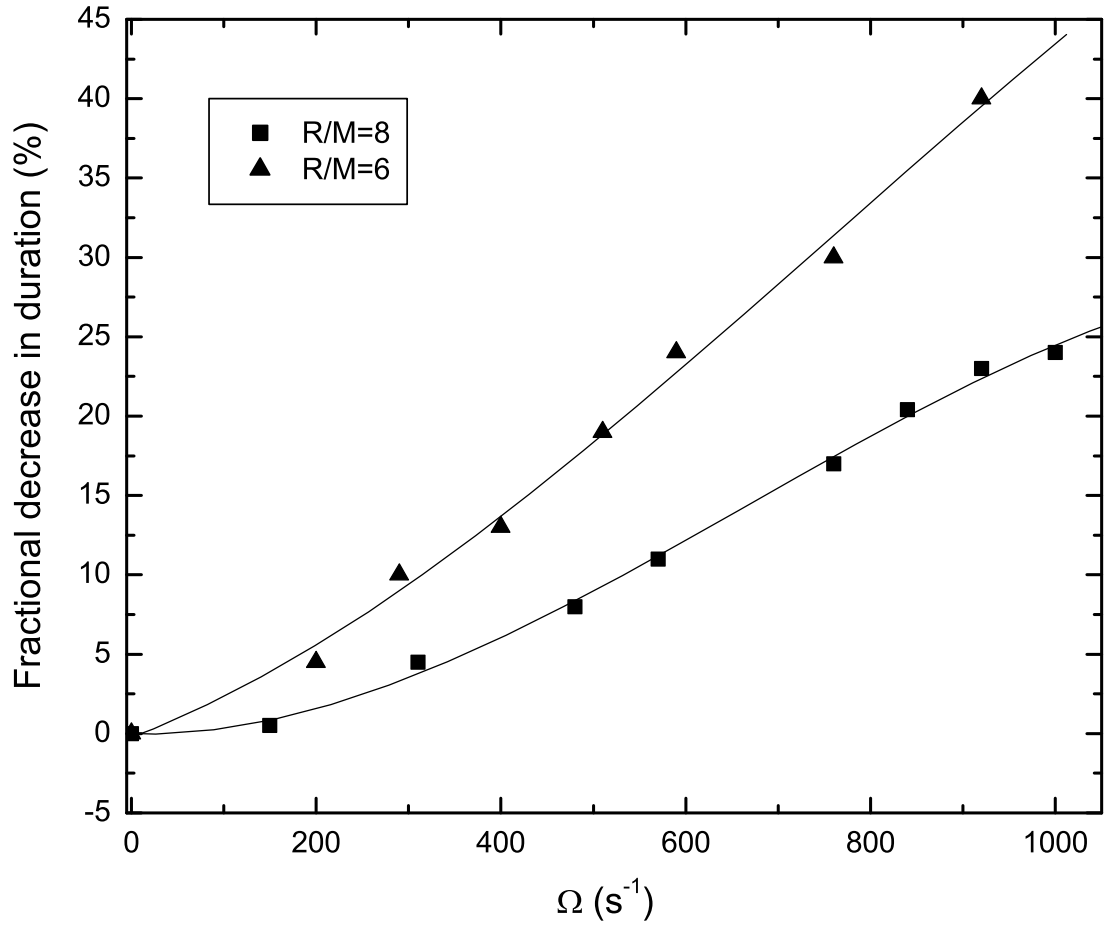}
\caption{Fractional decrease in the duration of thermal afterglow of `spot' case as a function of rotational frequency for $R/M\sim 8$ and $R/M\sim 6$.}
\end{figure}

\clearpage
\begin{figure}
\epsscale{0.7}\plotone{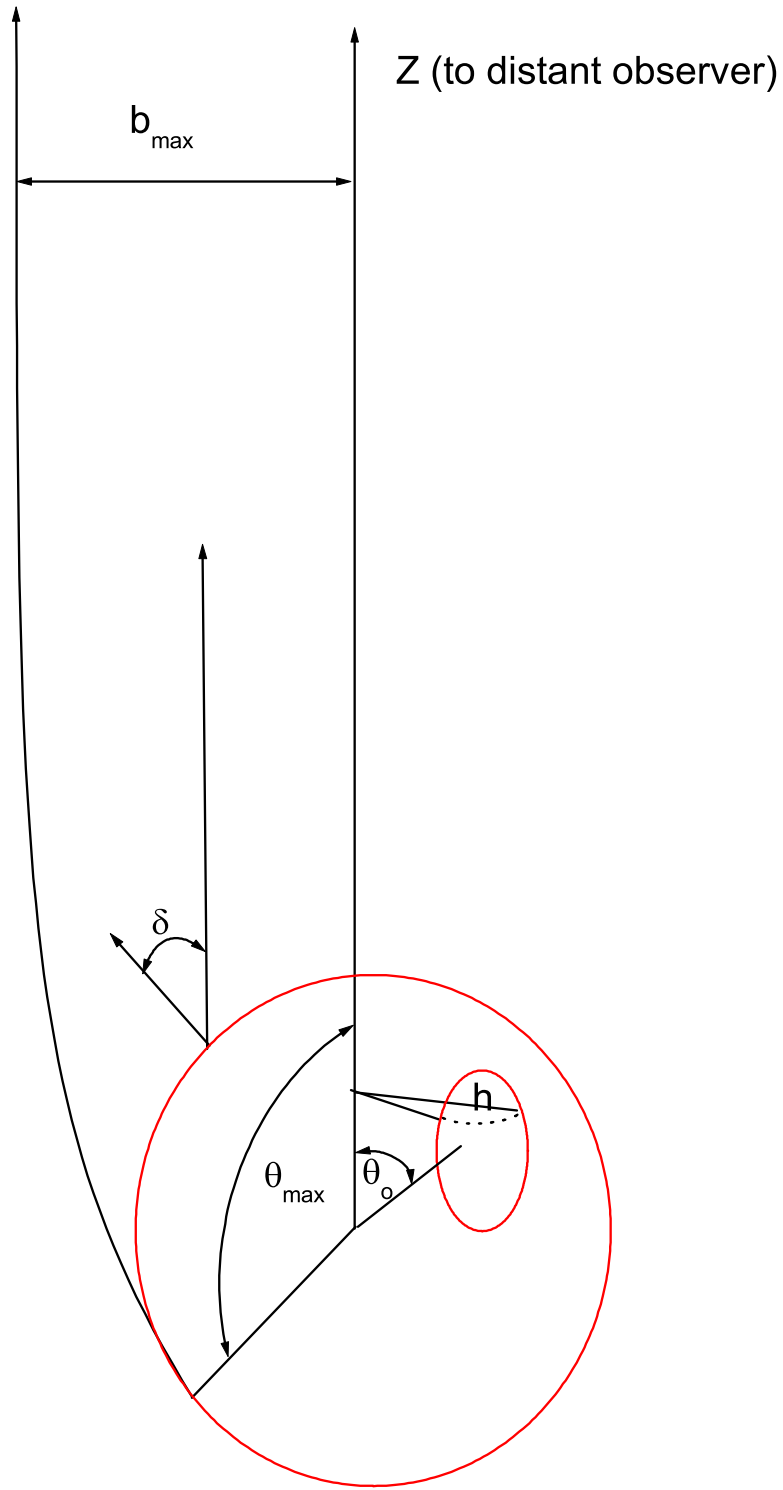}
\caption{The geometry used to determine the brightness of a hot spot.}
\end{figure}

\clearpage
\begin{figure}
\epsscale{1}\plotone{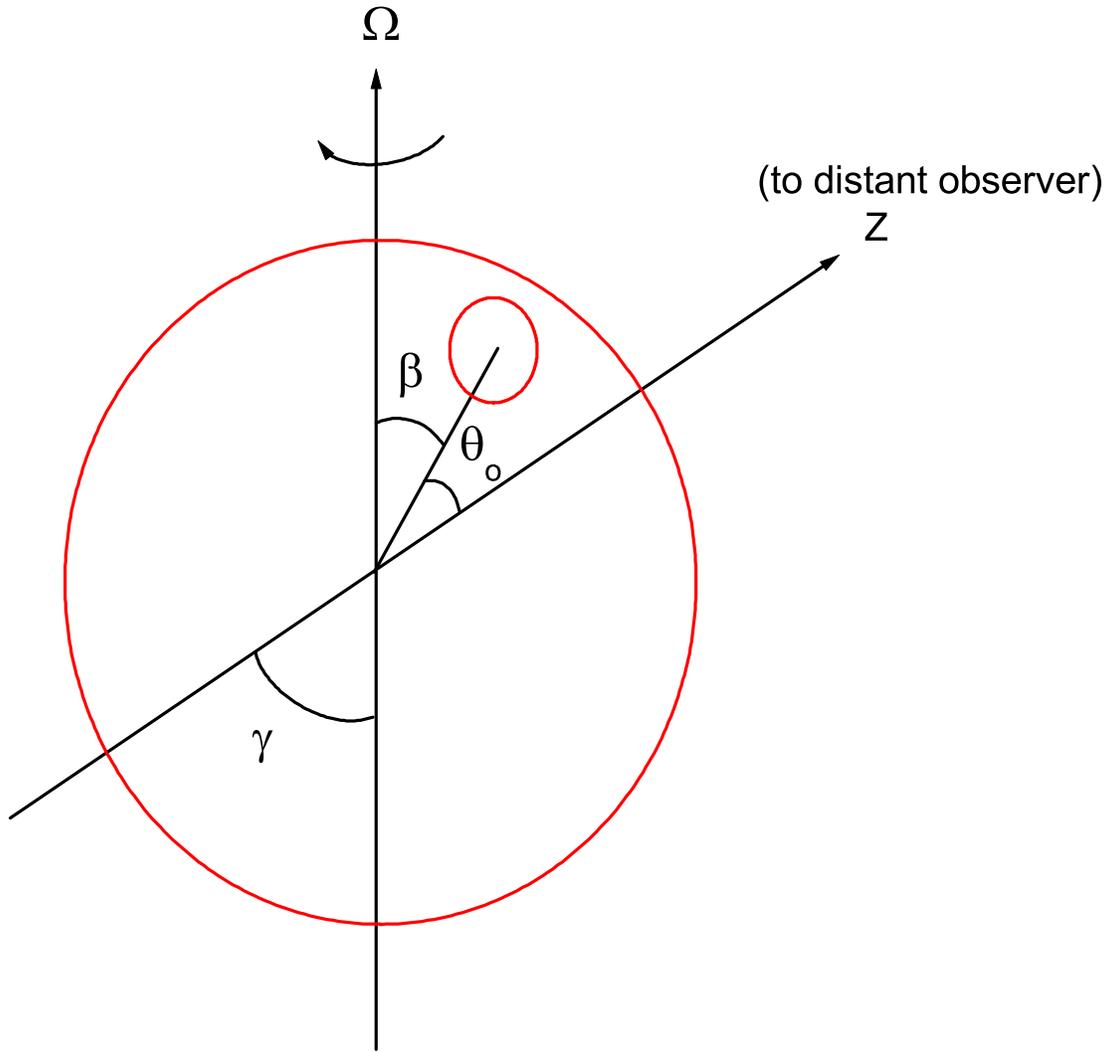}
\caption{The relation of the orientation of the hot spot and the axis of rotation to the observer's line of sight.}
\end{figure}

\clearpage
\begin{figure}
\plotone{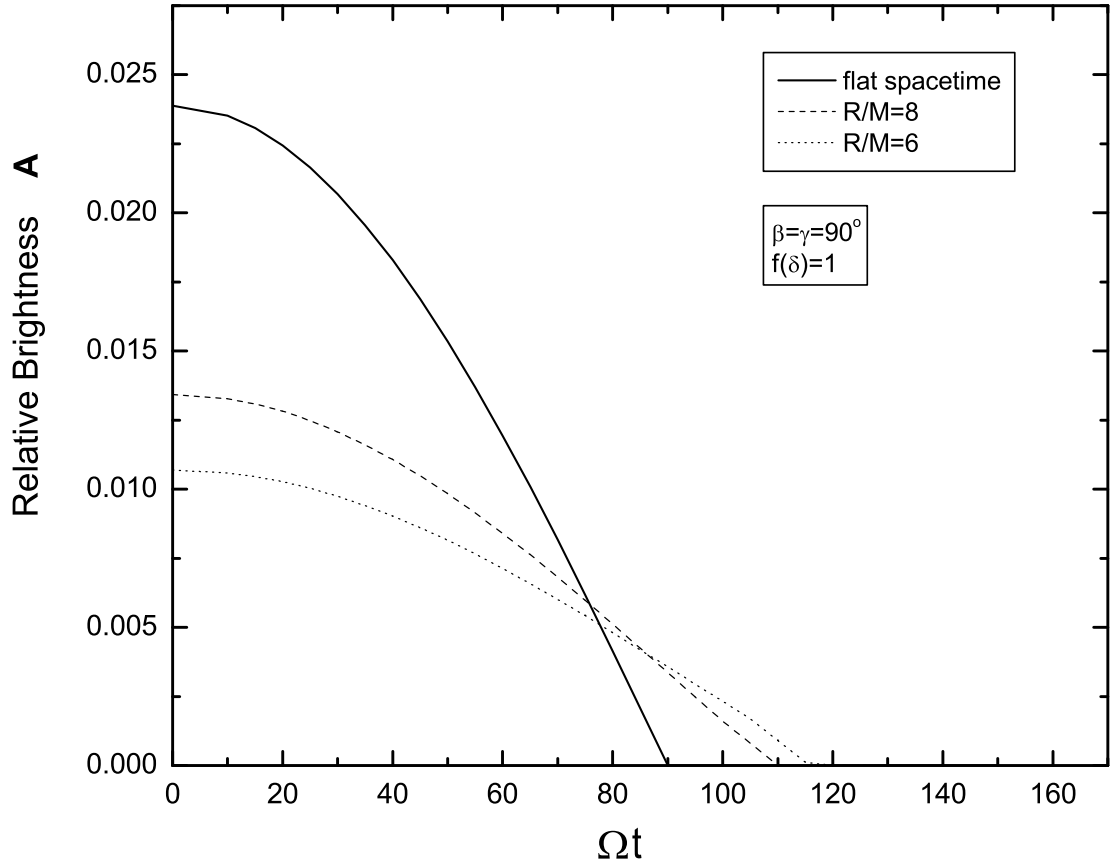}
\caption{Relative brightness $A$ as a function of phase for $\alpha=5^{\circ}$, with $f\left(\delta\right)=1$, $\beta=\gamma=90^{\circ}$.}
\end{figure}

\clearpage
\begin{figure}
\plotone{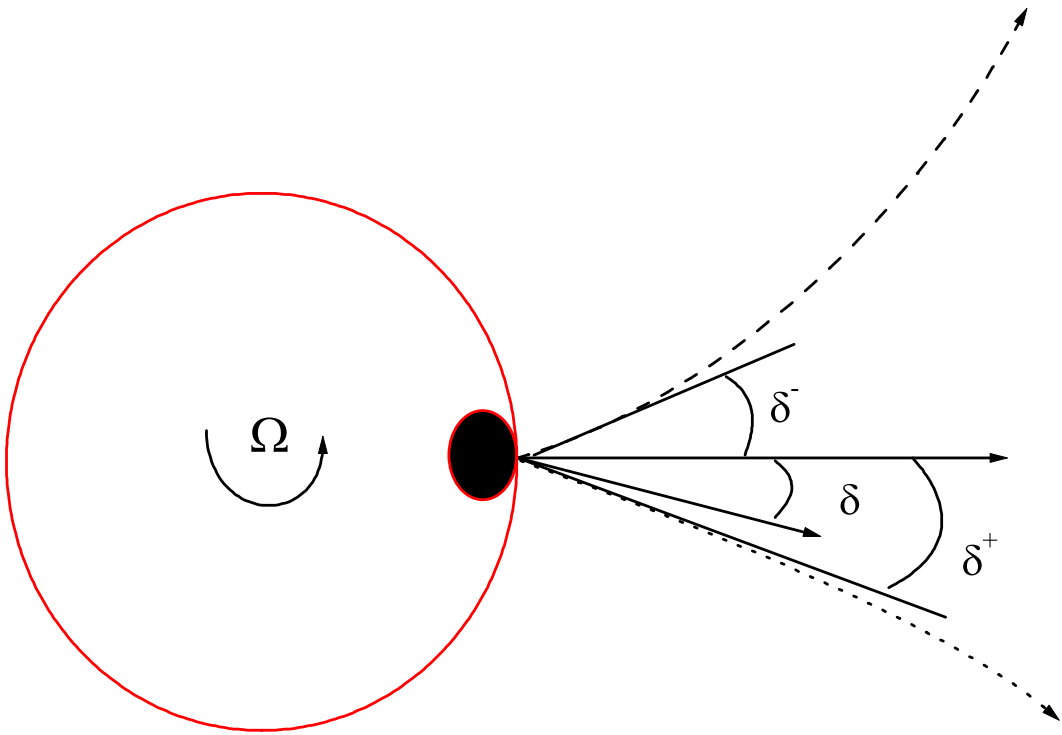}
\caption{Schematic illustration of the effect of frame dragging on the photon trajectory.}
\end{figure}

\clearpage
\begin{figure}
\plotone{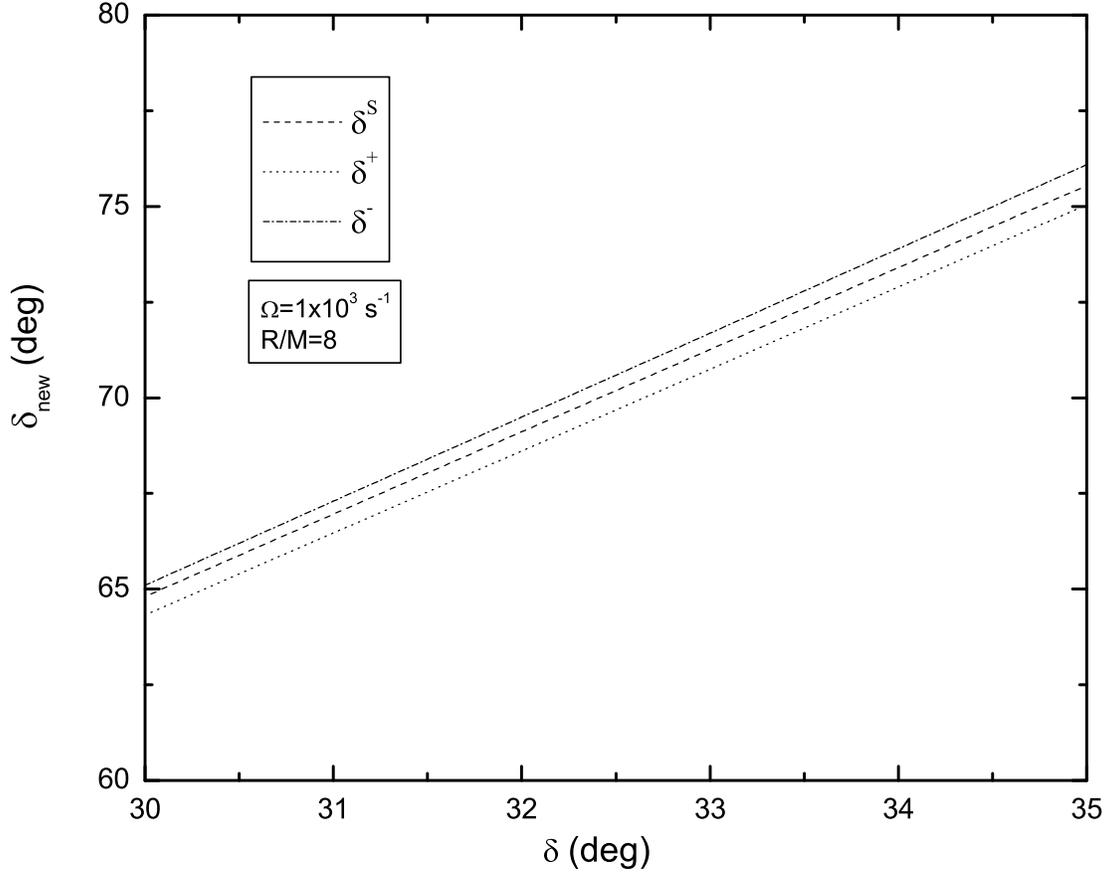}
\caption{Effect of frame dragging on the photon trajectory. $\delta^{S}$ illustrates the deflection in Schwarzschild spacetime and
$\delta^{+}, \delta^{-}$  illustrate the deflection in tangentially backward direction and tangentially forward direction at the rotational equator of a neutron star respectively.}
\end{figure}

\clearpage
\begin{figure}
\plotone{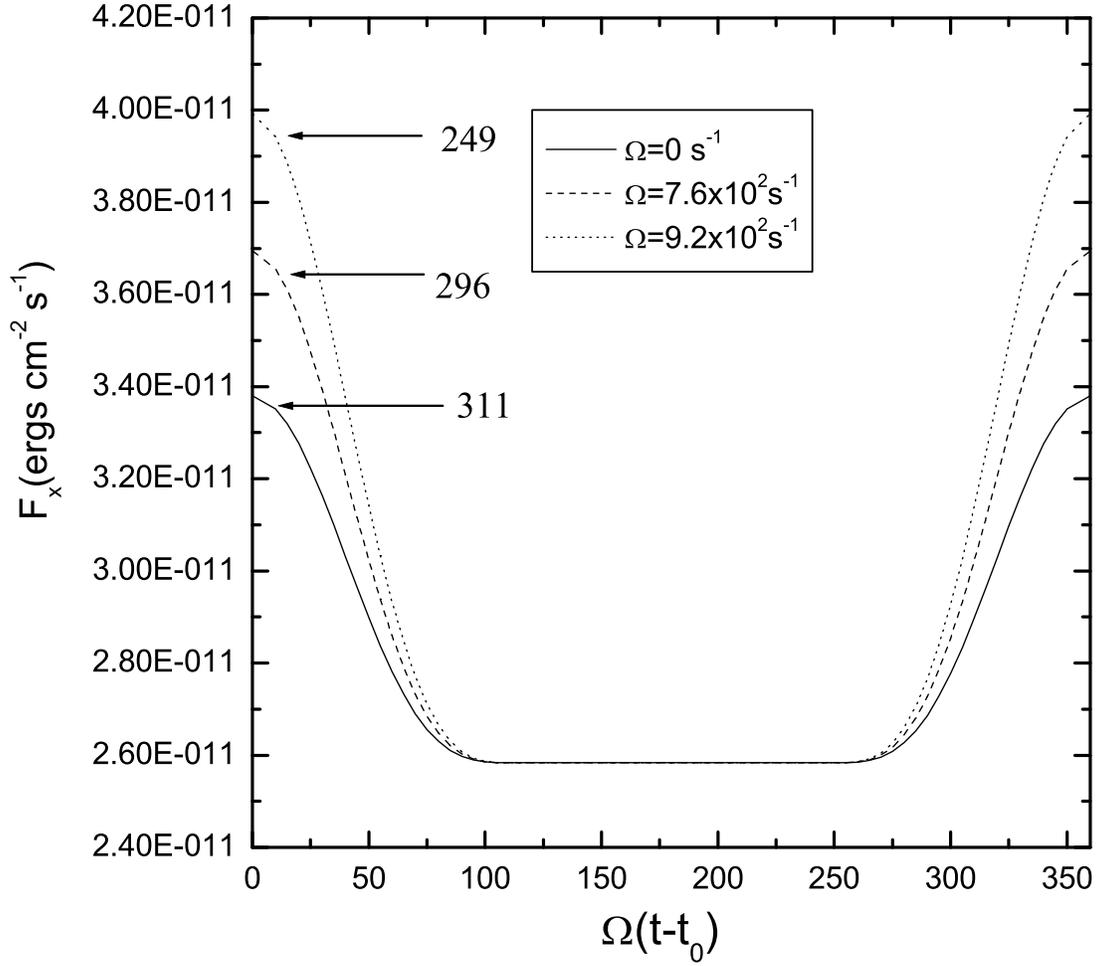}
\caption{Thermal X-ray pulse profiles at the peak time for a neutron star with $R/M\sim 8$, $T_{c}=10^{7} \mbox{K}$, $\Delta E=10^{42}\mbox{erg}$ at $\rho_{glitch}=10^{13}\mbox{g cm}^{-3}$, $\theta=\phi=90^{\circ}$ in an orientation with $\alpha=3^{\circ}$, $f\left(\delta\right)=\cos\delta$, $\beta=\gamma=90^{\circ}$ for `spot' case. Three cases with different rotational frequency are compared.
The numbers associated with the arrows indicate the number of day after the glitch}
\end{figure}

\clearpage
\begin{figure}
\plotone{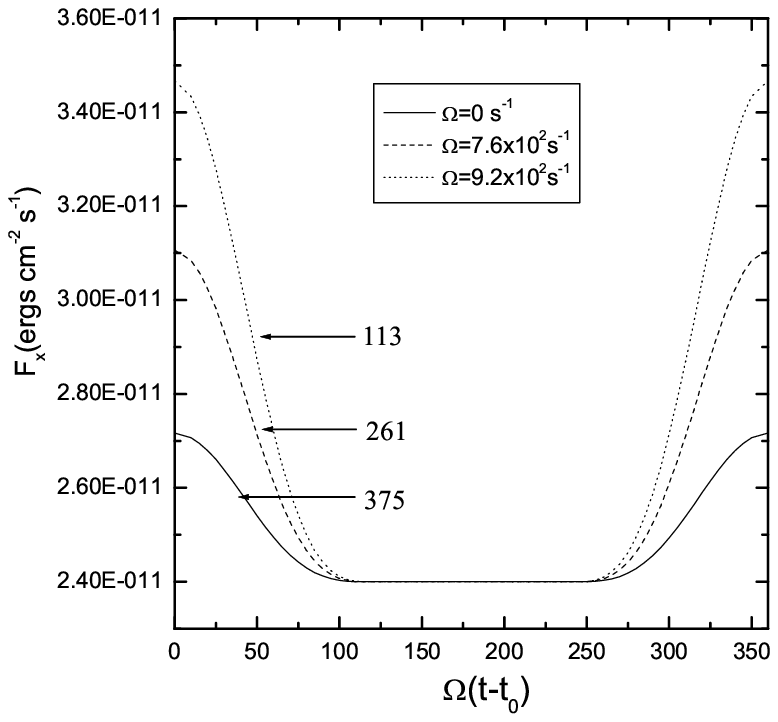}
\caption{Same as figure 11 but with $R/M\sim 6$.}
\end{figure}

\clearpage
\begin{table}
\begin{center}
\begin{tabular}{|rr|rr|}
\tableline
   $R/M=8$ &    &$R/M=6$ &  \\ 
\tableline\tableline
 $\Omega$ $s^{-1}$ &$ \Delta D_{r}$ $cm^{2}s^{-1}$ & $\Omega$ $s^{-1}$ & $\Delta D_{r}$ $cm^{2}s^{-1}$\\
\tableline
152 & 14.1 & 200 & 28.6\\
305 & 56.3 & 300 & 60.3\\
500 & 152.2 & 513 & 183.0\\
579 & 201.3 & 586 & 238.0\\
762 & 344.6 & 760 & 395.8\\
839 & 416.1 & 802 & 418.2\\
\tableline
\end{tabular}
\end{center}
\caption{The deviation of radial diffusion coefficient from the non-rotational value (i.e. $\Delta D_{r}=D_{r,\Omega\not= 0}-D_{r, \Omega =0}$) at the location of energy deposition for `spot' cases (i.e. at $\rho_{glitch}=10^{13} \mbox{g cm}^{-3}, \theta =\phi =90^{\circ}$ with $\Delta E=10^{42}\mbox{erg}, T_{c}=10^{7}\mbox{K}$) at $t=0~s$ as a function of rotational frequency.}
\end{table}

\end{document}